\documentclass[prb,twocolumn,aps,preprintnumbers,showpacs,tightenlines,floatfix,amssymb]{revtex4}
\usepackage{epsfig}

\newcommand{\beqn}{\begin{eqnarray}}
\newcommand{\eeqn}{\end{eqnarray}}
\newcommand{\eq}[1]{(\ref{#1})}
\newcommand{\ang}{\,\mbox{{\AA}}}
\newcommand{\cA}{{\mathcal A}}
\newcommand{\cB}{{\mathcal B}}
\newcommand{\dd}{{d}}

\begin{document}

\title{Spontaneous superconductivity and optical properties of high-$T_c$ cuprates}

\author{M.N.~Chernodub}
\affiliation{ITEP, B.Cheremushkinskaya 25, Moscow, 117218, Russia \\ and ITP, Kanazawa University, Kanazawa 920-1192, Japan}

\begin{abstract}
We argue that the high temperature superconductivity in cuprate compounds may
be supported by interaction between copper-oxygen layers mediated by {\it in-plane} plasmons.
The strength of the interaction is determined by the $c$-axis geometry and by the $ab$-plane
optical properties. Without making reference to any particular in-plane mechanism
of superconductivity, we show that the interlayer interaction favors spontaneous
appearance of the superconductivity in the layers.
At a qualitative level the model describes correctly the dependence of the transition temperature on
the interlayer distance, and on the number of adjacent layers in multilayered homologous
compounds. Moreover, the model has a potential
to explain (i) a mismatch between the optimal doping levels for critical temperature and
superconducting density and (ii) a universal scaling relation between the dc-conductivity, the
superfluid density, and the superconducting transition temperature.
\end{abstract}



\maketitle

The layered structure of the cuprates is a well established
fact\cite{ref:Anderson} related to a strong anisotropy in their electronic and optical
properties\cite{ref:Basov}. The common feature of all cuprates --- famous for their ability
to exhibit superconductivity at high transition temperatures ---
is the presence of conducting CuO${}_2$ layers separated by the so called charge reservoirs. These
reservoirs are nearly insulating even in superconducting phase.\cite{ref:Basov}

One of the most fascinating features of high temperature superconductivity is a strong dependence
of basic superconducting properties (including the most important quantity, the
transition temperature $T_c$) on the $c$-axis structure of cuprates. In particular, there is
a systematic dependence of the critical temperature on the number $n$ of the closely packed
CuO${}_2$ layers per a structural $c$-axis unit.
In typical homologous series of superconducting cuprates the
separation between the $n$ multilayers lies in the range $d \sim 6-15\ang$ which is large compared
to the spacing $c_{\mathrm{int}} \approx 3.5\ang$ separating individual layers inside
the multilayer. Since $c_{\mathrm{int}}$ is numerically close to the in-plane Cu-O bond length,
$a(\mbox{Cu-O}) \approx 3.8\ang$, it seems reasonable to treat the multilayer as a single thick layer.

Reflectance data indicate that the dielectric functions of the cuprates may qualitatively
be described by the so-called two-fluid model,\cite{ref:two:fluid,ref:Basov}
\beqn
\varepsilon(\omega) = \varepsilon_\infty - \frac{\omega^2_{ps}}{\omega (\omega + i 0^+)}
- \frac{\omega^2_{pn}}{\omega [\omega + i \gamma_{pn}(\omega)]}\,,
\label{eq:two:fluid}
\eeqn
which is a generalization of a simplest metallic dielectric function
$\varepsilon(\omega) = 1 - \omega^2_{\mathrm{p}}/[\omega (\omega + i \gamma_{\mathrm{p}})]$.
The carriers are divided into normal and superfluid components which have different
impacts onto optical and conducting properties of the cuprates.
In Eq.~\eq{eq:two:fluid} $\omega_{ps}$ and $\omega_{pn}$ are the plasma frequencies of
the superconducting and normal components, respectively, and
$\varepsilon_\infty$ is the high-frequency limit of $\varepsilon$. The relaxation processes of
the normal-state electrons are described by (in general, frequency-dependent)
scattering rate $\gamma_{pn}(\omega)$.
In ordinary metals the relaxation rate is small compared to the
plasma frequency ({i.e.},
$\gamma^{\mathrm{Al}}_{\mathrm{p}}/\omega^{\mathrm{Al}}_{\mathrm{p}} \approx 5\times 10^{-3}$ for aluminium).
Contrary to the metals,
{\it both} the finite conductivity and the relaxation processes are essential for
optical properties of the cuprates\cite{ref:Basov} because in a typical cuprate
$\gamma_{pn}(\omega_{pn}) / \omega_{pn}  \sim 1$.

Both in-plane and out-plane reflectance data show a sharp drop at the
frequencies higher than the plasma edge\cite{ref:Uchida} $\omega_{ps}/\sqrt{\varepsilon_\infty}$,
which determines a boundary of a transparency window. Despite existence of other descriptions of
the optical conductivities\cite{ref:Basov,ref:two:component} we take Eq.~\eq{eq:two:fluid} as
our starting formula for the sake of concreteness.

The form of the dielectric function~\eq{eq:two:fluid} suggests
the presence of plasmon-mediated phenomena at the energy scales
governed by the characteristic plasma frequencies. These phenomena
are usually studied with respect to the $c$-axis conductivity
(``transverse plasmon'').\cite{ref:Basov,ref:c:axis:plasmon}
The importance of the $ab$-plasmons for a proper description of the superconducting
state in layered materials such as high-Tc cuprates was clearly stressed
in Ref.~\onlinecite{ref:Andreas}. In our complimentary study we show that despite the $ab$-plane
plasmon is heavily damped\cite{ref:c:axis:plasmon} it induces the spontaneous appearance
of the superconductivity in the layers. Philosophically, our approach
resembles mechanisms based on the interlayer Josephson
tunneling\cite{Anderson:ITL} and interplane Coulomb
interaction,\cite{ref:Legett} as well as other
approaches\cite{ref:critique} including phenomenological models
of the Ginzburg-Landau type.\cite{ref:GL:like}

The free energy per one $d$ period per unit layer area $S$ in the absence of external
fields is given by a sum of the contributions from the normal ($F_n$) and the
superconducting ($F_s$) states of the layer, and the plasmon-mediated interaction
between the multilayers ($F_{\mathrm{pl}}$):
\beqn
F = F_n(\omega_{pn},\gamma_{pn}) + F_s(\omega_{ps}) + F_{\mathrm{pl}}(\omega_{ps},\omega_{pn},\gamma_{pn})\,.
\label{eq:FFF}
\eeqn
In each term we explicitly indicate the leading-order dependence on the optical
parameters $\omega_{ps}$, $\omega_{pn}$, and $\gamma_{pn}$. Long-range modulations of the
$c$-axis structure are neglected. We imply that the effect of the intralayer media is solely
insulating\cite{ref:Basov} thus neglecting a small finite out-plane conductivity in the normal state.

The free energy of the normal state $F_n$ in \eq{eq:FFF} should depend on the optical parameters
$\omega_{pn}$ and $\gamma_{pn}$ related to a specific (in fact, model-dependent) behavior of
electrons in individual CuO${}_2$ layers.
Since we would like to keep our approach as general as possible we exclude $F_n$ from our analysis concentrating on
the difference in the free energies of the normal and superconducting states,
\beqn
\delta F = F_s(\omega_{ps}) + F_{\mathrm{pl}}(\omega_{ps},\omega_{pn},\gamma_{pn}) - F_{\mathrm{pl}}(0,\omega_{pn},\gamma_{pn})\,.
\label{eq:d:F}
\eeqn

The free energy density of the superconducting state in Eq.~\eq{eq:FFF} is written
in the Ginzburg-Landau (GL) form\cite{ref:StatPhys}
\beqn
F_s = \! \frac{1}{S} \! \int_{V_l}\!
\Bigl[\frac{\hbar^2}{4 m^*} {\Bigl|\Bigl({\vec\nabla} - \frac{2 i e}{\hbar c} {\vec A}\Bigr)\psi\Bigr|}^2
\!\!+ {\cA} {|\psi|}^2 + \frac{{\cB}}{2} {|\psi|}^4\Bigr] \dd V,
\label{eq:F:s:1}
\eeqn
where the integration is going over the volume $V_l$ of the superconducting layer,
$2 m^*$ is the effective mass of the superconducting carrier, and ${\cA}$, ${\cB}$ are the
GL phenomenological parameters describing the behavior of the order parameter $\psi$
which is related to the density of the condensed electrons $n_s = |\psi|^2$.
The GL approach has known limitations,
while being usually correct near the point of the superconducting transition.
Universality arguments suggest that the GL parameters
must depend on the intrinsic layer properties
while being generally less dependent on the intralayer structure.

Under assumption of a spatial
homogeneity of the order parameter $\psi$ and negligence of fluctuations of the electromagnetic field
${\vec A}$, the supercurrent in~\eq{eq:F:s:1} vanishes and we
arrive to the simple expression
\beqn
F_s(\omega_{ps}) = w_n \cdot \Bigl({\cA}\eta\, \omega_{ps}^2 + \frac{{\cB} \eta^2}{2} \omega_{ps}^4 \Bigr)\,,
\quad \eta = \frac{m^*}{16 \pi^3 e^2}\,,
\label{eq:F:s:2}
\eeqn
where $w_n$ is the multilayer width. We used the relations
\beqn
{|\psi|}^2 \equiv n_s = \frac{m^* c^2}{4 \pi e^2 \lambda_L^2} \equiv \eta\, \omega_{ps}^2\,, \qquad
\lambda_L \omega_{ps} = 2 \pi c\,,
\label{eq:psi}
\eeqn
where $\lambda_L$ is the London penetration depth.

Naively, if the layers were structureless very thick solid plates made of alike atoms
interacting with the van der Waals potential $U(r) = - \kappa\, r^{-6}$, then the
interaction energy of the layers would be described by the well-known Hamaker
form\cite{ref:StatPhys},
$U_{\mathrm{pl}}(d) = - H/(12 \pi d^2)$, where $H = \kappa \pi^2 \rho^2$
is the Hamaker constant and $\rho$ is the number density of atoms in the planes.

\begin{figure}
\begin{center}
\includegraphics[scale=0.85,clip=true]{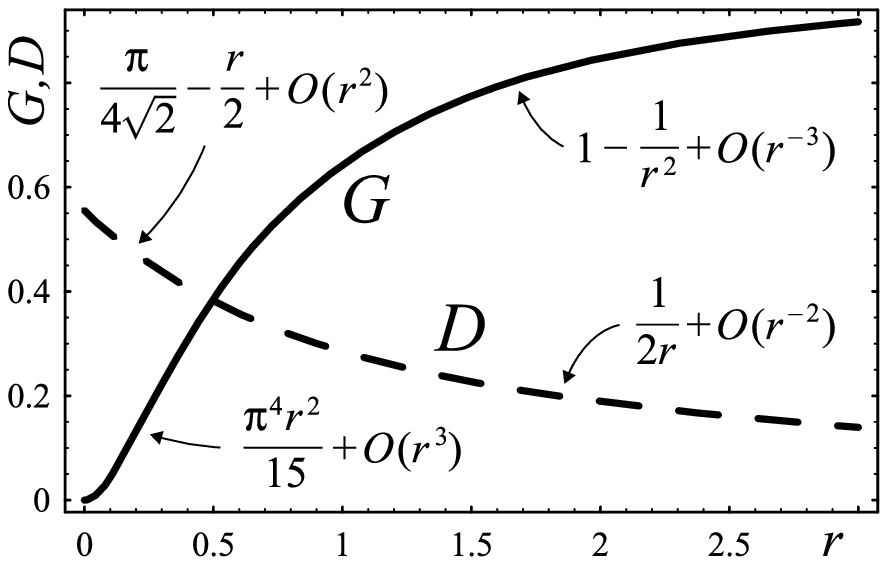}
\end{center}
\caption{The suppression factors: geometrical, $G$, Eq.~\eq{eq:G},
and dissipative, $D$, Eq.~\eq{eq:R}, and their asymptotics at $r\to0,\infty$. }
\label{fig:G}
\end{figure}

None of the above assumptions is satisfied by the cuprate layers because of
significance of retardation, relaxation, dielectric absorbtion and geometrical
suppression effects. These effects are known\cite{ref:StatPhys} to diminish
the interaction which still follows the Hamaker law $U_{\mathrm{pl}} \propto d^{-2}$.
Up to an inessential numerical coefficient
\beqn
U_{\mathrm{pl}} = - \hbar \Omega\, \frac{G(w_n/d)}{16 \pi^2 \epsilon_{\mathrm{int}} \, d^2}\,, \quad
\Omega = \int_0^\infty {\left[\frac{\varepsilon(i \xi) - 1}{\varepsilon(i \xi) + 1}\right]}^2 \dd \xi\,,
\label{eq:Upl}
\eeqn
where $\epsilon_{\mathrm{int}}$ is the intralayer dielectric function.
The geometrical factor $G$ takes into account the ``multilayer-insulator'' periodic
structure~\footnote{For simplicity we treat all layers equally while in general
({i.e.}, for $n \geqslant 2$ Hg-based series) this is not the case.}
\beqn
G(r) = \frac{1}{(1+r)^2} \Bigl[\psi^{(1)}\Bigl(\frac{1}{1+r}\Bigr) + \psi^{(1)}\Bigl(\frac{1+2r}{1+r}\Bigr)
- \frac{\pi^2}{3}\Bigr],
\label{eq:G}
\eeqn
where $\psi^{(1)}$ is the first derivative of the digamma function. The geometric factor $G$ (shown
in Fig.~\ref{fig:G} by the solid line) is a monotonically increasing function of $r \equiv w_n/d$.

The interaction energy~\eq{eq:Upl} is of an electromagnetic origin. In an idealized limit
of perfectly conducting layers the interaction energy may be imagined as the Casimir energy\cite{ref:Casimir}
of the electromagnetic field stored between the layers\cite{ref:StatPhys}. In the case of real materials
we follow Ref.\onlinecite{ref:surface:plasmon} and interpret Eq.~\eq{eq:Upl} as the interaction energy
between the layers caused by the inlayer plasmons. The plasmons give a dominant contribution to the interlayer energy at short
interlayer separations\cite{ref:surface:plasmon} $\omega_{pn} \, d \ll 1$. This condition is satisfied by typical cuprates
({e.g.}, $\omega_{pn} \, d \lesssim 10^{-3}$ for the La${}_{2-x}$Sr${}_x$CuO${}_4$ compound discussed below). The
characteristic frequency\cite{ref:StatPhys} $\Omega$ in Eq.~\eq{eq:Upl} gives account of the absorbtion spectra
of the layers which, in turn, characterize the strength of the interaction between the $ab$-plasmons.

The frequency $\Omega$ can be expressed via the dielectric function $\varepsilon(\omega)$ evaluated at the
imaginary axis $\omega = i \xi$. The dis\-per\-sion relation\cite{ref:StatPhys} expresses
$\varepsilon(i \xi)$ via the conductivity $\mbox{Re}\,\sigma = \omega \, \mbox{Im}\,\varepsilon/(4\pi)$ at the real axis:
\beqn
\varepsilon(i \xi) = 1 + \frac{2}{\pi} \int\limits_0^\infty\! \frac{\omega \, \mbox{Im}\, \varepsilon(\omega)}{\omega^2 + \xi^2} \dd \omega
\equiv 1 + 8 \int\limits_0^\infty \frac{\mbox{Re}\, \sigma(\omega)}{\omega^2 + \xi^2} \dd \omega.
\label{eq:dispersion}
\eeqn
Thus, the interlayer interaction~\eq{eq:Upl} is fixed by the dissipative part of the in-plane
conductivity, $\mbox{Re}\,\sigma$, the interlayer dielectric parameter $\varepsilon_{\mathrm{int}}$,
and the $c$-axis geometry.

Despite the scattering rate $\gamma_{pn}$ in cuprates is a complicated
frequency-dependent function\cite{ref:Basov} one can approximately evaluate the order
of the characteristic frequency $\Omega$ in the normal state ($\omega_{ps} = 0$)
assuming that $\gamma_{pn}$ is $\omega$-independent: $\Omega = \omega_{pn} \, D(\gamma_{pn}/\omega_{pn})$.
The dissipative suppression factor [derived from Eqs.~\eq{eq:two:fluid}, \eq{eq:Upl}, \eq{eq:dispersion}],
\beqn
D(r) = \frac{r \sqrt{r^2-2} - 4 {\mathrm{arcsinh}} \sqrt{(r/\sqrt{2}-1)/2}}{2 {(r^2-2)}^{3/2}}\,,
\label{eq:R}
\eeqn
is a monotonically decreasing function of $r \equiv \gamma_{pn}/\omega_{pn}$. We plot
$D(r)$ in Fig.~\ref{fig:G} by the dashed line.

To estimate the energy scales related to the plas\-mon-me\-di\-ated interactions, we
consider La${}_{2-x}$Sr${}_x$CuO${}_4$ (La214) compound. In La214 the CuO${}_2$ layers
are perfectly flat and are separated by two LaO layers at the distance $d=c_0/ 2\approx 6.6\ang$.
The basic cell has the tetragonal structure $a_0 \times b_0 \times c_0$ with the base
parameters $a_0 \approx b_0 \approx 3.8\ang$. The condensation energy of the optimally doped
($x=0.16$, $T_c = 38\,\mathrm{K}$) La214 compound is known to be
${\mathcal E}^{\mathrm{(Cu)}}_{\mathrm{cond}} \approx 13\,\mu\mbox{eV}$ per one atom of copper.\cite{ref:La214:condensation}
The normal state of the slightly underdoped
La214 is characterized\cite{ref:La214:Gao} by the plasma frequency $\omega_{pn} \sim 6000\,\mbox{cm}^{-1}$ while
the typical scattering rate is of the order $\gamma_{pn} \sim 2000\,\mbox{cm}^{-1}$ for frequencies higher than the
$ab$-plane ``pseudogap'' $\omega_{ab} \approx 700\,\mbox{cm}^{-1}$.\cite{ref:La214:Startseva,ref:Uchida2}
Fixing the scattering rate to be constant and taking into account the
dissipative suppression factor, $D(1/3) \approx 0.4$, we get
$\Omega_{\mathrm{La214}} \approx 2500\,\mbox{cm}^{-1}$.
The characteristic frequency is of the order of a typical superconducting gap
$\Delta \sim 10-50 \,\mbox{meV}$ in
the cuprates~\footnote{We ignore finite temperature corrections to the interlayer interaction~\eq{eq:Upl}
because $T_c \ll \hbar \Omega_{\mathrm{La214}}/k_B \approx 600\,\mbox{K}$.},
$\hbar \, \Omega_{\mathrm{La214}} \sim 50\,\mbox{meV}$.

In order to avoid suspicious fine-tuning of parameters in Eq.~\eq{eq:Upl} we roughly set
$\varepsilon_{\mathrm{int}} \sim 1$, $c_0 \sim d$ (then $a_0 \sim d$ and the geometrical suppression is
$G(1) \approx 0.6$). Then the plasmon-mediated interaction energy per copper atom is
\beqn
U^{\mathrm{(Cu)}}_{\mathrm{pl}}({\mathrm{La214}}) =
- \frac{G\hbar \, \Omega_{\mathrm{La214}}}{16 \pi^2 \epsilon_{\mathrm{int}}} {\Bigl(\frac{a_0}{d}\Bigr)}^2
\approx - 200\,\mu{\mathrm{eV}}\,.
\eeqn
This value is by an order of magnitude higher than the condensation energy
$U^{\mathrm{(Cu)}}_{\mathrm{pl}} \sim 10 \, {\mathcal E}^{\mathrm{(Cu)}}_{\mathrm{cond}}$.
In other words, the condensation energy may well be explained by a $10\%$ deviation in the interplane interaction, which
in turn, should be related to a change of similar magnitude in the optical parameters
of a cuprate as it cools down from critical to lower temperatures. In fact, the optical characteristics of
cuprate compounds vary essentially in this range\cite{ref:Basov} exhibiting, {e.g.},
a sharp drop of the scattering rate $\gamma_{ps}$ and dominance of the
superconducting component $\omega_{ps}$ at $T<T_c$.
This argument stresses importance of the $ab$-plasmon mediated inter-layer interactions.
Below we ignore all interlayer interactions except for Eq.~\eq{eq:Upl}.

\begin{figure}
\begin{center}
\includegraphics[scale=0.85,clip=true]{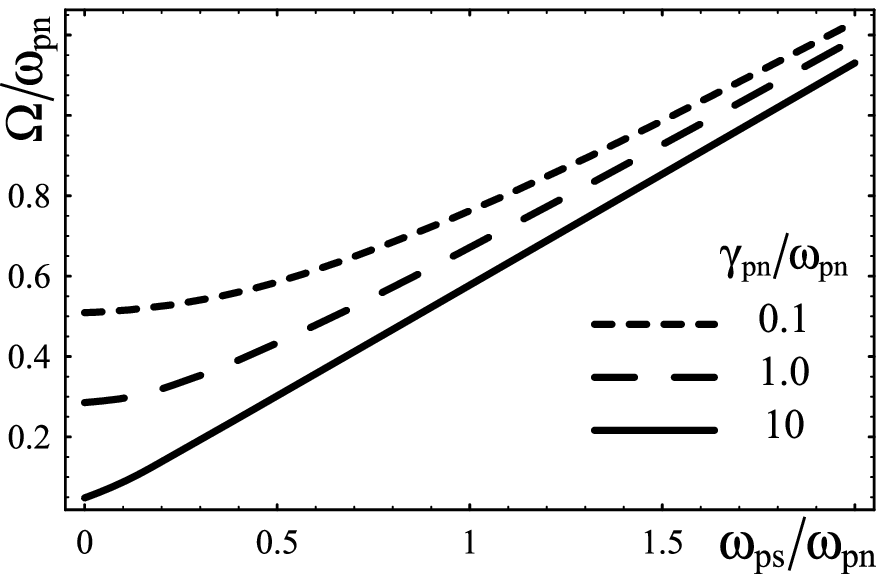}
\end{center}
\caption{The characteristic frequency~\eq{eq:Upl} $\Omega$
{\it vs} the superconducting frequency $\omega_{ps}$
for various scattering rates $\gamma_{pn}$.}
\label{fig:omega}
\end{figure}

The crucial property of the interlayer interaction term $U_{\mathrm{pl}}$ is that
it favors appearance of the scatterless superconducting component with $\omega_{ps} \neq 0$.
To illustrate this property we expand the characteristic frequency $\Omega$, Eq.~\eq{eq:Upl},
at $T = T_c$ in powers of $\omega_{ps}/\omega_{pn} \ll 1$,
\beqn
\Omega(\omega_{ps}) = \omega_{pn} \Bigl\{u_0 + u_2 \cdot {\Bigl(\frac{\omega_{ps}}{\omega_{pn}}\Bigr)}^2
+ O\Bigl[{\Bigl(\frac{\omega_{ps}}{\omega_{pn}}\Bigr)}^4\Bigr]\Bigr\},
\label{eq:Omega:exp}
\eeqn
The dimensionless coefficients $u_m$ are certain functionals of the scattering rate
${\bar \gamma}_{pn}(y) = \gamma_{pn}(y\cdot \omega_{pn})/\omega_{pn}$, {e.g.}
\beqn
u_2[{\bar \gamma}_{pn}] = \int_0^\infty \frac{4 [y+{\bar \gamma}_{pn}(y)]^2 \, \dd y}{\{2y[y+{\bar \gamma}_{pn}(y)]+1\}^3}
\label{eq:u2}\,.
\eeqn
As one can see from Eq.~\eq{eq:u2} the second coefficient of the
expansion~\eq{eq:Omega:exp} is {\it always} positive, $u_2>0$, regardless of particularities of the scattering
rate $\gamma_{pn}(\omega)$. The behavior of $\Omega$ at $\gamma_{pn}=\mbox{const}$ is illustrated in Fig.~\ref{fig:omega}.

Since the characteristic frequency $\Omega$ enters Eq.~\eq{eq:Upl} with the minus sign,
the interaction energy~\eq{eq:Upl} may provoke a tachyonic instability against emergence
of the superconducting condensate~\eq{eq:psi}, $|\psi|^2 \propto \omega_{ps}^2$.
In other words, the interlayer interaction~\eq{eq:Upl} supports the
appearance of a superconducting ($\psi \neq 0$) state provided the layers are
intrinsically able --- via {\it any} microscopic mechanism --- to generate this superconducting state.

At high temperatures the energy density associated with the superconducting condensate~\eq{eq:F:s:1}, \eq{eq:F:s:2}
is higher compared to the gain in free energy~\eq{eq:d:F} which would be achieved by the plasmon interaction~\eq{eq:Upl}.
This makes the superconductivity energetically unfavorable.
As the temperature decreases the coefficient ${\cA}(T)$ in the GL free energy~\eq{eq:F:s:2} gets gradually smaller
and at the certain temperature the quadratic $\omega^2_{ps}$ term of the GL free energy~\eq{eq:F:s:2} cancels
the same term in the plasmon-mediated interaction~\eq{eq:Upl}, \eq{eq:Omega:exp}. This cancelation marks the
critical temperature $T_c$. At lower temperatures, $T<T_c$, the overall coefficient in front of the quadratic
term turns negative and the system becomes unstable against spontaneous development of the superconductivity
$\omega_{ps} \sim |\psi| \neq 0$.

The relation between the critical temperature ($T_c$), optical ($\gamma_{pn}$, $\omega_{pn}$) and
geometrical ($w_n$, $d$) parameters of the cuprates can be derived from Eqs.~\eq{eq:d:F}, \eq{eq:Upl}, \eq{eq:Omega:exp}:
\beqn
{\cA}(T_c) \cdot \frac{\omega_{pn} m^* \varepsilon_{\mathrm{int}}}{\pi \hbar \, e^2\, u_2(\gamma_{pn}/\omega_{pn})}
= \frac{1}{w_n\, d^2}\, G\bigl(w_n/d)\,,
\label{eq:omega:pn}
\eeqn
where the left hand side (LHS) contains optical and microscopic parameters
at $T=T_c$ while the right hand side (RHS) is of the purely geometrical origin. Below we list
a few universal features of the cuprate superconductors which are described by Eq.~\eq{eq:omega:pn}.

{\it Transition temperature $T_c$ vs $n$.}
The GL coefficient ${\cal A}(T)$ is a monotonically increasing function of temperature.
Thus, the higher (lower) value of the RHS in \eq{eq:omega:pn}, the higher (lower) value of $T_c$
is~\footnote{We assume that the normal-state parameters are weakly dependent on temperature around $T_c$.}.
The RHS of Eq.~\eq{eq:omega:pn} is a linearly-rising function of $w_n$ at $w_n \ll d$.
At $w_n = 0.4448\, d$ the RHS has a maximum and then it decreases as $1/w_n$ at
$w_n \gg d$ (we used Eq.\eq{eq:G} as well as the asymptotics of $G$ given in Fig.~\ref{fig:G}).
In $n$-layered cuprates the width of the multilayer is a monotonically rising function of $n$, which can approximately
be estimated as $w_n = (n-1) (c_{\mathrm{int}} + \delta_g) + \delta_c$, where
$\delta_g$ is the geometrical width of a single layer, ranging from
$\delta_g(\mbox{La-214}) = 0\ang$ and $\delta_g(\mbox{Bi-2212})=0.013\ang$ to
$\delta_g(\mbox{YBCO})=0.274\ang$,\cite{ref:StructuralCommonalities}
$c_{\mathrm{int}} \approx 3.5\ang$ is the interlayer spacing inside the multilayer,
and $\delta_c$ is a ``coherence width'' of external layers which should be of the order of
the $c$-axis coherence length $\xi_c$ (a few~$\ang$).
For typical crystallographic parameters the RHS of \eq{eq:omega:pn}, and, consequently, the
transition temperature $T_c$, are peaked around $n_{\max}=3$.
This behavior is in fact a universal feature of the homologous series.\cite{ref:Anderson}

{\it Transition temperature $T_c$ vs $d$.} The {\it r.h.s} of~\eq{eq:omega:pn} is a monotonically
decreasing function of the separation $d$ between the multilayers provided the other geometrical
parameters of the $c$-axis structure are fixed. Thus, the larger $d$ the lower temperature must be.
This is another universal behavior observed in the cuprates.\cite{ref:Uemura:review}

{\it Transition temperature $T_c$ vs $x$.} One may expect that the highest $T_c$
is achieved at the doping $x$ at which the density $n_s$ of the superconducting
carriers is highest. However, this expectation is not confirmed experimentally:\cite{ref:optimality} the optimal doping
for the transition temperature is noticeably lower compared to the one for the carriers ({i.e.}, in
La-214, Y-123, Bi-2212 cuprates one has $x^{T_c}_{\mathrm{opt}}\approx 0.16 < x^{n_c}_{\mathrm{opt}} \approx 0.19$).
The plasmon-mediated interaction may explain this behavior. If the RHS of Eq.~\eq{eq:omega:pn} were independent
of $x$ then the maximum temperature would be achieved at a certain value of $T=T_c(x^{n_s}_{\mathrm{opt}})$ corresponding to
the highest carrier density. However, the interlayer distance $d$ increases with the doping $x$,\cite{ref:doping:c:dependence}
lowering the plasmon interaction energy (proportional to the RHS of \eq{eq:omega:pn}). Thus, the equality~\eq{eq:omega:pn}
is achieved at a lower value of the GL parameter, ${\cA}[T_c(x^{n_s}_{\mathrm{opt}})]<{\cA}[T_c(x^{T_c}_{\mathrm{opt}})]$,
implying $T_c(x^{n_s}_{\mathrm{opt}}) < T_c(x^{T_c}_{\mathrm{opt}})$.

{\it Scaling between $T_c$, $\omega_{ps}$ and dc-conductivity}.
At sufficiently low temperatures the normal component is almost invisible in
the dielectric function.\cite{ref:Basov,ref:La214:Startseva}
In this case $\Omega = \pi \omega_{ps}/(4 \sqrt{2})$
[we used \eq{eq:R} as well as the $\gamma_{pn} \to 0$ asymptotic, Fig.~\ref{fig:G}],
and \eq{eq:Upl} becomes linear in $\omega_{ps}$:
\beqn
U_{\mathrm{pl}}(\omega_{ps}, T=0) = - \frac{G(w_n/d)}{64 \sqrt{2} \pi \epsilon_{\mathrm{int}} \, d^2}\, \hbar \omega_{ps}(0)\,.
\label{eq:Upl:T0}
\eeqn
Neglecting the quartic term in \eq{eq:F:s:2} we get the superconducting frequency at $T=0$ as a minimum of \eq{eq:d:F}:
\beqn
\omega_{ps}(0)\cA(0) = \frac{\pi\cA(T_c)}{8 \sqrt{2}}
u^{-1}_2\Bigl(\frac{\gamma_{pn}(T_c,\omega)}{\omega_{pn}(T_c)}\Bigr)\cdot \omega_{pn}(T_c)\,,
\label{eq:relation:omega:0}
\eeqn
where we used Eq.~\eq{eq:omega:pn} and disregarded the variation of the crystallographic parameters in the range of
temperatures between $T_c$ and $T=0$. The relation~\eq{eq:relation:omega:0} links the ratio of the GL layer's
parameter $\cA$ at $T=T_c$ and $T=0$ with both superconducting and normal optical properties of the
$ab$-planes. Note that (i) the LHS (RHS) of Eq.~\eq{eq:relation:omega:0} depends solely on $T=0$ ($T=T_c$) quantities;
(ii) the relation~\eq{eq:relation:omega:0} is {\it universal}: it is does not depend on the $c$-axis structure and should
hold for all cuprate materials with the same in-plane parameters.

Since the relation~\eq{eq:relation:omega:0} is dependent on $\omega_{pn}(T_c,\omega)$
further analytical calculations are difficult. We notice, however, that the integral~\eq{eq:u2}
is saturated at low frequencies relevant to the normal-state dc-conductivity:
\beqn
\sigma_{{\mathrm{dc}}} \equiv \lim_{\omega\to 0} \mbox{Re}\,\sigma(\omega) = \frac{\omega^2_{pn}(T_c)}{4 \pi}
\, \lim_{\omega\to 0} \gamma_{pn}^{-1}(T_c,\omega)\,.
\label{eq:sigma:dc}
\eeqn
Therefore we take $\gamma_{pn}$ to be equal to its low-frequency extrapolation, and work in
the ``dirty'' limit  $\gamma_{pn} \gg \omega_{pn}$, arriving to $u_2 = \gamma_{pn}/\omega_{pn}$.
The requirement of the dirty limit is rather mild since even at $\gamma_{pn} = \omega_{pn}$
the above relation holds within $10\%$. Next, we take the standard
GL-like prescription $\cA(T) = \alpha (T+{\widetilde T}_{c})$, with $0 < {\widetilde T}_{c} \ll T_c$,
where $\alpha$ and ${\widetilde T}_{c}$ are the GL parameters describing the free energy associated
with the condensation. We consider the most energetically unfavorable case:
the positive ${\widetilde T}_c$ indicates that the layers alone are not able
to support the superconductivity. Curiously, if the intrinsic layer properties are related,
${\widetilde T}_{c} = \beta \, \omega_{ps}(0)$ with $\beta \approx 10 \, \mbox{K}\cdot\mbox{cm}$, then
from \eq{eq:relation:omega:0} and \eq{eq:sigma:dc} we get the scaling relation
$\omega^2_{ps}(0) \approx 120\, \sigma_{\mathrm{dc}} \, T_c$ observed experimentally.\cite{ref:universal:Basov}

\begin{acknowledgments}
The author is thankful to the members of Institute for Theoretical
Physics of Kanazawa University and especially to Prof. Tsuneo Suzuki
for the kind hospitality and stimulating environment.
The work was supported by the JSPS Grant No. L-06514.
\end{acknowledgments}

\end{document}